\documentclass[12pt]{revtex4-1}
\usepackage{amssymb,graphicx,xspace,color}
\setcitestyle{super}

\usepackage{natbib}

\begin{document}

\textbf{\LARGE{Strain-derivative of thermoelectric properties: a sensitive probe for nematicity}}\\
\\
\\F. Caglieris$^{1}$, C. Wuttke$^{1}$, X. C. Hong$^{1}$, S. Sykora$^{1}$, R. Kappenberger$^{1}$, S. Aswartham$^{1}$, S. Wurmehl$^{1}$, B. B\"{u}chner$^{1,2,3}$ \& C. Hess$^{1,3}$\\
\\
$^{1}$\textit{Leibniz-Institute for Solid State and Materials Research, 01069 Dresden, Germany}\\
$^{2}$\textit{Institut f\"{u}r Festk\"{o}rperphysik, TU Dresden, 01069 Dresden, Germany}\\
$^{3}$\textit{Center for Transport and Devices, TU Dresden, 01069 Dresden, Germany}\\
\\
\\
\\

\textbf{The nematic instability of iron-based superconductors is an undebatable ingredient of the physics of iron-based superconductors. Yet, its origin remains enigmatic as it involves a fermiology with an intricate interplay of lattice-, orbital- and spin- degrees of freedom. It is well known that thermoelectric transport is an excellent probe for revealing even subtle signatures of instabilities and pertinent fluctuations. In this paper, we report a strong response of the thermoelectric transport properties of two underdoped 1111 iron-based superconductors to a vanishingly small strain. By introducing the strain-derivative of the Seebeck and the Nernst coefficients, we provide a novel description of the nematic order parameter, proving the existence of an anisotropic Peltier-tensor beside an anisotropic conductivity-tensor. Our measurements reveal that the transport nematic phenomenology is the result of the combined effect of both an anisotropic scattering time and Fermi surface distortions, pointing out that in a realistic description, abreast of the spin-fluctuations also the orbital character is a fundamental ingredient. In addition, we show that nematic fluctuations universally relax in a Curie-Weiss fashion above $T_S$ in all the elasto-transport measurements and we provide evidences that nematicity must be band-selective.}\\

%\end{abstract}
\textbf{Introduction}

The investigation of nematic orders in solid state systems has been strongly boosted in recent times by the hypothesis of their intimate link with the emerging unconventional superconductivity in copper-based \cite{Daou:2010aa,Lawler:2010aa} and specially in iron-based superconductors \cite{Fernandes:2014aa,Chuang181}. In the latter, the nematic order identifies a lowering of the rotational symmetry characterized by a tetragonal-to-orthorhombic structural transition, which typically anticipates the formation of a magnetic order with additional signatures of orbital ordering \cite{Fernandes:2014aa,Chuang181}. This crossover of multiple orders generate a "chicken-or-egg" problem, whose solution is considered a crucial step towards the understanding of the unconventional superconductivity  \cite{Fernandes:2014aa}. 
One of the most successful experimental approach to this dilemma is the study of the strain-suceptibility of different physical quantities in proximity of the structural transition. Remarkably, the use of the strain derivative of the electrical resistivity as a sensitive quantity mimic of the nematic susceptibility allowed to track the evolution of the nematic fluctuations and to distinguish their electronic origin from a simple ferroelastic distortion \cite{Chu710,Kuo958,Hosoi8139}. Although this promotes the nematic order to the same level of the other electronic instabilities (i.e. superconductivity and density-waves), its microscopic origin remains a puzzle, in particular concerning the role of the orbital and spin degrees of freedom. 

It is common wisdom that thermoelectric transport properties are excellent probes to investigate the fermiology of unconventional materials. In the vicinity of electronic instabilities, they could exhibit spectacular effects, resulting in an extreme sensitivity to phase transitions and fluctuations of order parameters \cite{0953-8984-21-11-113101,0953-2048-29-7-073002, PhysRevB.83.224505, PhysRevB.83.092507, 0034-4885-79-4-046502}. Therefore, in proximity of an electronic nematic instability, the evaluation of the thermoelectric response to an applied uniaxial strain, appears as an ideal experimental approach to unveil the role of Fermi Surface distortions.
In this regard, particularly promising is the observation of a strong anisotropy in the Seebeck ($S$) and the Nernst ($\nu$) coefficients in the nematic phase of some detwinned 122-iron based superconductors \cite{FUJII201631, PhysRevLett.110.067001,PhysRevB.97.100506,PhysRevB.97.220501}, as a direct indication of a possible strong elasto-thermoelectric effect.\
%in this work we propose a new modus operandi to investigate the response of  a system when approaching a nematic instability.
%Other intriguing approaches include both the effects, such as the orbital-selective spin-nematic model, which takes into account the orbital character of the spin fluctuations \cite{PhysRevB.97.121109, PhysRevB.99.155117, PhysRevB.94.155138}. 
%In this sense, the description of the nematic order parameter provided by the elasto-resistivity, results incomplete.
In order to exploit this potential, we introduce two novel quantities, namely the elasto-Seebeck effect and the elasto-Nernst effect, defined as the strain-derivative of $S$ and $\nu$ in the limit of vanishingly small strain $\epsilon$. Experimentally, we take advantage of an innovative setup, which combines a standard thermoelectric measurement configuration, with the highly controlled uniaxial strain offered by a piezoelectric device. By measuring directly the slope of $S$ and $\nu$ vs $\epsilon$, we have access to the respective strain derivatives (Figure 1, see Materials and Methods for details on the experimental setup), representing the nematic susceptibility of the system. In particular, in this work, we investigate the transport nematic phenomenology of two single crystals of LaFe$_{1-x}$Co$_x$AsO with x=0 and 0.035 \cite{KAPPENBERGER20189,WANG201950}, respectively. The former is the parent compound and the latter an electron underdoped sample of the La-1111 family, so far almost unexplored due to the lack of sizeable single crystals.\\
We will show that the thermoelectric transport properties react extraordinarily to a tiny strain well above the nematic transition, revealing an extended zone of nematic fluctuations with a diverging behavior at the structural transition. Moreover, by analyzing the contribution of the different transport coefficients we demonstrate that a band-selective response of the Peltier tensor $\bar \alpha$ is indispensable to quantitatively explain the behavior of the thermoelectric coefficient under strain, pointing out the fundamental role of the Fermi surface distortions caused by electron-nematic order \cite{PhysRevB.80.220514}. \
In addition, we experimentally demonstrate that, within the validity of a single-band approximation, the elasto-Nernst coefficient can be predicted as a combination of the strain-derivatives of the Peltier $\bar \alpha$ and the conductivity $\bar \sigma$ tensors. This lets us paint a self-consistent scenario in which all the elasto-transport coefficients are bound to each other.\\

%In particular, in this work, we investigate the transport nematic phenomenology of two single crystals of LaFe$_{1-x}$Co$_x$AsO with x=0 and 0.035 \cite{KAPPENBERGER20189,WANG201950}, respectively. The former is the parent compound and the latter an electron underdoped sample of the La-1111 family, so far almost unexplored due to the lack of sizeable single crystals.\\
 
%The study of this family is indeed crucial because it presents the highest critical temperatures \cite{Paglione:2010aa} and differently from other families of iron-based superconductors, such as the 122, the 1111 exhibits a large temperature gap between the structural and the magnetic transition \cite{Hess1111, WANG201950}.

%In contrast to previous reports on thermoelectric properties \cite{FUJII201631, PhysRevLett.110.067001,PhysRevB.97.100506,PhysRevB.97.220501}, we obtained an analogous divergent behavior for $T>T_S$ also in a crystal of BaFe$_2$As$_2$ \cite{SM}. The qualitative difference between those and our own measurements is explainable in terms of different experimental conditions, namely small strain-limit vs large finite stress.\\  

%Figure 1 shows the different steps of our measurement procedure, described in details in the 'Materials and Methods' section. However, it is interesting to notice in the top panels of Figures 1b and 1e that the response of the Seebeck and the Nernst signals ($\Delta V_S$ and $\Delta V_{\nu}$ to the applied strain is surprisingly large, considering that we operate in the linear regime of the low-strain ($\sim 0.0001 - 0.001$) limit (Figure 1c).

\textbf{Elasto-Seebeck}\\

Figures 2a and 2b show the temperature-dependence of Seebeck coefficient $S$ of the x=0 and the x=0.035 sample, respectively. Their amplitude and trend are consistent with previous reports on polycrystalline compounds \cite{0953-2048-29-7-073002}. In the x=0 sample a sign change of $S$ occurs at around 170 K, caused by the multi-band nature of this material. Since hole-like and electron-like pockets contribute to the Seebeck coefficient with opposite sign, they tend to compensate their respective effect, possibly generating a change in the sign of $S$, as in our case.
In the x=0.035 compound, the electron doping, obtained by Co-substitution, pushes it closer to the condition of single-carrier transport and the Seebeck coefficient remains always negative, as expected for a system dominated by electrons. This has been already shown in La(Fe,Co)AsO and Sm(Fe,Co)AsO series of polycrystalline samples, where a departure from the carrier compensation in favour of an electron-like transport due to Co-doping has been demonstrated \cite{PhysRevB.79.054521, doi:10.1063/1.4766936}. 

Figures 2c and 2d present the temperature dependence of the strain ($\epsilon$) derivative of the Seebeck coefficient $\delta(\Delta S/T)/\delta \epsilon$, with $\Delta S$=$S$($\epsilon$)-$S$($\epsilon$=0), for the x=0 and x=0.035 compound, respectively. The normalization to $T$ is introduced to get rid of the entropy contribution. Interestingly, in both the compounds a change of regime appears in correspondence of the structural transition, whose onset is around $T_S$=155 K in the parent compound and $T_S$=80 K for the underdoped one \cite{WANG201950}. While $\delta(\Delta S/T)/\delta \epsilon$ of the x=0 compound exhibits a sharp cusp-like transition, in the x=0.035 compound the change of regime at $T_S$ looks like a crossover. However, for $T>T_S$, where the crystalline cell is tetragonal and nematic fluctuation are expected, $\delta(\Delta S/T)/\delta \epsilon$ is finite and large in both the compounds, evidencing a strong response of the Seebeck coefficient to a vanishingly small strain. Moreover, $\delta(\Delta S/T)/\delta \epsilon$ exhibits a diverging trend by approaching $T_S$. This behavior is typically detected in the context of elasto-resistivity measurements, where a Curie-Weiss fashion of the elasto-resistivity is interpreted as the fingerprint of large nematic fluctuations with an electronic origin \cite{Chu710,Kuo958,XC-arxiv}.\ 

In addition, with our experimental approach we could also take advantage of the fact that $S$ is sensitive to the sign of the charge carriers. Indeed, the most striking result of our measurement concerns the sign of $\delta(\Delta S/T)/\delta \epsilon$: In the parent compound, though $S$ changes sign, $\delta(\Delta S/T)/\delta \epsilon$ is always positive and it remains finite also when $S$ crosses the zero. This points out that not all the different Fermi pockets are responsible for the electronic nematic phenomenology but only some of them contribute. Interestingly, the strain derivative persists to be positive also in the x=0.035 sample, in which the transport properties are dominated by the electron-like carriers. This suggests that if the nematicity is band-selective, the electron-pockets play the major role.\\

\textbf{Elasto-Nernst}\\

Figure 3a and 3d show the temperature-dependence of the Nernst coefficient $\nu$ of the x=0 and the x=0.035 sample, respectively. First of all, one can notice that, differently from the Seebeck coefficient, $\nu$ is positive in the considered temperature range for both the compounds. Indeed, in the Nernst effect the contribution of different types of carriers is independent on the sign of their electric charge \cite{PhysRevLett.91.066602}. Hence, the contributions by hole-like and electron-like pockets tends to sum up instead of compensating (it is also called ambipolar Nernst effect \cite{PhysRevLett.91.066602}). This is consistent with the decrease of the absolute value of the Nernst coefficient from the x=0 to the x=0.035 compound, interpreted as a departure from the perfect bi-polarity caused by the electronic doping. In addition, in the parent compound, $\nu$ undergoes an abrupt increase at $T_S$, consistent with previous reports \cite{0953-2048-29-7-073002, PhysRevB.83.092507}. This has been sometimes attributed to the band reorganization caused by the development of the long range magnetic ordering and the consequent appearance of Dirac-cone-like bands \cite{0953-2048-29-7-073002}.\

Figures 3b and 3e present the temperature dependence of the strain derivative of the Nernst coefficient $\delta(\Delta \nu/T)/\delta \epsilon$ as a function of $T$, where $\Delta \nu$=$\nu$($\epsilon$)-$\nu$($\epsilon$=0). In analogy to the strain-derivative of the Seebeck effect, also $\delta(\Delta \nu/T)/\delta \epsilon$ presents a diverging behavior in the tetragonal phase and a change of regime at around $T_S$. Differently from the Seebeck coefficient, the ambipolar nature of the Nernst effect prevents $\nu$ to cross the zero value in our compounds. This allows to normalize $\delta(\Delta \nu/T)/\delta \epsilon$ to the value of the unstrained $\nu$ without incurring the risk of an unphysical divergency and define a Nernst nematic susceptibility $\chi_\nu$=$\delta(\Delta \nu/\nu)/\delta \epsilon$. $\chi_\nu$ can be directly compared to the susceptibility calculated from the elasto-resistivity $\chi_\rho$=$-\delta(\Delta \rho/\rho)/\delta \epsilon$ \cite{XC-arxiv}, whose value is generally assumed as the response of the electronic nematic order parameter to the applied strain \cite{Chu710,Kuo958,XC-arxiv}. The result is presented in Figures 3c and 3f for the x=0 and the x=0.035 sample, respectively. For $T>T_S$ our data are consistent with a Curie-Weiss fit (blue lines in Figures 3c and 3f) $\chi_\nu$=$\chi_0+C/(T-T^*)$, where $\chi_0$ is an intrinsic piezoelectric effect unrelated to the electronic nematicity, $C$ is the Curie constant and $T^*$ is the Curie temperature, corresponding to $T^*$=129$\pm$8 K and $T^*$=22$\pm$8 K for the x=0 and the x=0.035 sample, respectively. These values are in good agreement with the $T^*$ obtained by fitting $\chi_\rho$ \cite{XC-arxiv} (orange lines in Figures 3c and 3f) and it is generally interpreted as the mean field electronic nematic critical temperature \cite{Chu710,Kuo958}. This observation demonstrates the Nernst effect as a primary probe for detecting and tracking the electronic nematic fluctuations.\

Let's now focus on the absolute value of $\chi_\nu$ and $\chi_\rho$. Interestingly, $\chi_\nu$ tends to decrease from x=0 to x=0.035, reaching the respective maximal value of around 230 and 120 close to $T_S$. On the contrary $\chi_\rho$ increases with doping. The increase of $\chi_\rho$ due to Co-doping in the Ba(Fe,Co)$_2$As$_2$ series was attributed to an increase of the nematic fluctuation intensity towards the optimally-doped composition, which maximizes the superconducting critical temperature \cite{Chu710,Kuo958}. By assuming that both $\chi_\nu$ and $\chi_\rho$ should reflect the response of the nematic order parameter to the applied strain, the reason for their mismatch deserves some consideration. 
 The Nernst coefficient is a complex quantity which results from a non-trivial combination of of the resistivity tensor $\bar \rho$ and the Peltier thermoelectric tensor $\bar \alpha$, reading $\nu$=$\alpha_{xy}\rho_{yy}-\alpha_{xx}\rho_{xy}$ \cite{PhysRevB.98.155116}. In this notation $x$ is the direction along which strain and heat gradient (or electric current) are applied, while $y$ is the transverse direction. In complex materials, such as the iron based superconductors, the experimental prediction of $\nu$ is usually unsuccessful due to the complications of the multi-band nature. However, in the next section, we show that the behavior of the elasto-Nernst of the x=0.035 compound (closer to the single band condition thanks to the doping) can be reasonably obtained from the other transport coefficients, in a self-consistent scenario.\\
  
%Usually the experimental verification of this relation and the prediction of the Nernst coefficient behavior is a tricky operation, which becomes extremely complicated if one must take into account the implications of a multi-band scenario. However, if the single band approximation can be reasonably applied to describe the transport properties, one can expect to obtain at least a qualitative agreement. In the next section, we show that the behavior of the elasto-Nernst of the x=0.035 compound (closer to the single band condition thanks to the doping) can be reasonably predicted from the other transport coefficients.\\

\textbf{Analysis of the transport coefficients of the x=0.035 compound}\\

First of all, in a single-band approximation, it is possible to evaluate also a Seebeck susceptibility $\chi_S$=$\delta(\Delta S/S)/\delta \epsilon$, shown in Figure 4a as a function of $T$. For $T>T_S$, we interpolated the $\chi_S$ curve  with a Curie-Weiss function. The fitting parameter $T^*$=25$\pm$8 is in very good agreement with the values obtained by fitting $\chi_\nu$ and $\chi_\rho$. The Seebeck coefficient is explicited in terms of transport coefficients as $S$=$\alpha_{xx}\rho_{xx}$, where $\alpha_{xx}$ and $\rho_{xx}$ are the diagonal terms of the Peltier and the resistivity tensors, respectively. It is immediate to verify that $\chi_S$=$\chi_\rho$+$\chi_{\alpha}$, where $\chi_{\alpha}$=$\delta(\Delta \alpha_{xx}/\alpha_{xx})/\delta \epsilon$. From these relations, one can evaluate the temperature dependence of $\alpha_{xx}$ and $\chi_{\alpha}$, presented in Figure 4b. It must be noticed that $\chi_{\alpha_{xx}}$ is of the same order of magnitude as $\chi_\rho$. However, they exhibit an opposite sign, which is understandable, considering that $\alpha_{xx}\sim d \sigma_{xx}/dE=-\rho_{xx}^{-1}d\rho_{xx}/dE$, where $\sigma_{xx}$ is the electrical conductivity and $E$ is the energy. 
%Moreover, though both $\chi_S$ and $\chi_\nu$ exhibit a maximum around the structural transition, the transport coefficients $\chi_\rho$ and $\chi_\alpha$ continue increasing below $T_S$. Hence, the change of regime observed in $\chi_S$ and $\chi_\nu$ is caused by a balance between $\chi_\rho$ and $\chi_\alpha$, rather than an abrupt change of their anisotropy at $T_S$. 
At this point, one can consider $\chi_\nu$ and verify wether it is experimentally obtainable by a combination of the other transport coefficients. In the limit of small strain, in which we operate, one can safely state that $\rho_{xx} \simeq \rho_{yy}$ and $\delta \rho_{yy}/\delta \epsilon$=$(1/Y_P)\delta \rho_{xx}/\delta \epsilon$, where $Y_P$ is the Poisson ratio of the piezoelectric device. The off-diagonal terms do not directly contribute to the anisotropy \cite{PhysRevB.80.220514}. Hence, we can evaluate $\delta( \nu)/\delta \epsilon$=$\alpha_{xy}\delta \rho_{xx}/\delta \epsilon - \rho_{xy}\delta \alpha_{xx}/\delta \epsilon$, where $\alpha_{xy}$ is obtained by combining the transport properties in unstrained conditions \cite{PhysRevB.98.155116}. The result is shown in Figure 4c in comparison with the experimental value. The two curves are in good agreement, demonstrating experimentally the validity of the description of transport properties in terms of conductivity and Peltier tensors. Several remarks are in order. First of all we point out that a self-consistent interconnection of transport coefficients is rarely proven experimentally. One might speculate that the clean result is facilitated by the use of strain-derivatives, which render spurious effects less important. Second, our analysis shows the necessity of two different contributing transport coefficients ($\rho_{xx}$ and $\alpha_{xx}$) to explain the behavior of $\nu$ under strain.\\ 
The validation of this method is an important result because it can be applied to the investigation of all the single-band systems in which the proximity to an instability or a phase transition renders electronic properties susceptible to the strain. This includes many cuprates superconductors and doped iron based superconductors.\\

\textbf{Discussion and conclusions}\\

Once established the existence of a finite $\chi_\alpha$ beside a finite $\chi_\rho$, one can conjecture on the microscopic mechanisms that determine the transport nematic phenomenology. 
%Generally, the anisotropy in transport properties can be caused by either an anisotropy in the in-plane scattering time or an anisotropy of the Fermi surface. 
Generally, a pure spin-nematic scenario mainly supports an anisotropic scattering time as a source for transport anisotropy \cite{PhysRevLett.107.217002, PhysRevLett.105.157003, PhysRevB.90.121104, PhysRevLett.113.127001}, while the pure orbital-ordering description takes mainly into account anisotropies of the Fermi surface parameters, such as the Fermi velocity \cite{PhysRevB.84.024528,PhysRevLett.103.267001,PhysRevLett.105.207202, PhysRevB.82.100504}. 

%respectively linked more to the spin or the orbital degree of freedom. It is common wisdom that thermoelectric properties are indeed very sensitive to details of the Fermi surface.
In the context of cuprate superconductors, it was explicitly predicted that the Nernst effect anisotropy is a very sensitive probe of Fermi surface distortions caused by electron-nematic order \cite{PhysRevB.80.220514}. This is caused by a large anisotropy in $\alpha$, which overcomes the anisotropy of $\rho$ and results particularly enhanced in correspondence of a change of the Fermi surface topology \cite{PhysRevB.80.220514}. This means that, if only an orbital anisotropy dominates the transport, a $\chi_\alpha$ substantially larger than $\chi_\rho$ can be expected. Since in our case $\left | \chi_\rho \right | \ge \left | \chi_\alpha \right |$, it is likely that a significant contribution from an anisotropic scattering time must be present.\\
%Hence, being aware that a realistic theoretical description of the specific material is fundamental to draw a complete scenario, our method allows to extract experimentally the behavior under strain of different transport coefficients. \\ 
On the other hand, it has been reported that the NMR spin-lattice relaxation rate $(T_1T)^{-1}$ of LaFeAsO is finite well above $T_S$, as a signature of persistent spin fluctuations \cite{PhysRevB.97.180405}. For $T<T_S$, $(T_1T)^{-1}$ increases a lot, before diverging in correspondence of $T_N$, where spin fluctuations freeze \cite{PhysRevB.97.180405}. Interestingly, neither $\chi_\rho$ nor $\chi_\nu$ seem to be sensitive to the magnetic transition and they do not follow the trend of $(T_1T)^{-1}$.  This suggests that they are not mimicking the spin susceptibility of the system. As a consequence, the existence of an anisotropic scattering time, directly linked to anisotropic spin fluctuations, is not sufficient to explain the transport anisotropy, but an orbital contribution from the distortion of Fermi surface must be included.
Hence, to shed light on the cryptic nematic phase of iron-based superconductors, it is evident that a theoretical picture which includes different microscopic mechanism must be adopted. In this sense the orbital-selective spin-nematic model is a promising candidate, since it predicts that, once the orbital character of the spin-fluctuations is taken into account, both the anisotropy in scattering rate and in the Fermi surface  parameters (i.e. the Fermi velocities) must play a substantial role \cite{PhysRevB.97.121109, PhysRevB.99.155117, PhysRevB.94.155138}.\\

In summary, we measured for the first time the strain-derivative of the Seebeck and the Nernst effect of two single crystals belonging to the 1111 family of iron-based superconductors. We observed that thermoelectric properties, in proximity of a nematic instability, are strongly susceptible to a vanishingly small strain. The inspection of the Seebeck effect provided a clear signature of the band-selective character of the nematic phenomenology in case of a multi-band system, which is a fundamental information for the definition of a nematic order parameter. 
In addition, by defining a Nernst nematic susceptibility, we experimentally demonstrated that an anisotropy in the resistivity tensor $\bar \rho$ is not enough to explain the behavior of the thermoelectric properties, but a finite anisotropy in the Peltier tensor $\bar \alpha$ must be included. This suggests that the transport nematic phenomenology is likely to be the result of the combined effect of both an anisotropic scattering time and Fermi surface distortions, pointing out that in a realistic description, beside the spin-fluctuations also the orbital character is a fundamental ingredient.
We expect that these results will trigger novel theoretical insights, setting new bounds for the anisotropic transport models and giving a substantial contribution to the understanding of the nematic puzzle.\newpage

\textbf{Materials and Methods}

\textbf{Crystal growth}

The crystals were obtained using the solid-state single crystal growth method at ambient pressure using Na-As as a liquid phase promoting an abnormal grain growth due to enhanced interfacial anisotropy by introducing a liquid-solid interface. This is a different strategy from the usually used flux growth. As this growth is based on polycrystalline starting materials, a polycrystalline sample of LaFeAsO was prepared using a two-step solid-state reaction. The obtained polycrystalline pellets and Na-As powder were layered into an alumina crucible. The molar ratio of LaFeAsO to Na-As used was 1:4, which corresponds to a ratio in volume of about 1:1. The material was heated to 1080$^\circ$ C and annealed for 200 h. By using this method single crystals sized up to 2x3x0.4 mm$^3$ were obtained. Ref. \cite{KAPPENBERGER20189} gives a detailed description of the synthesis process of all the investigated compounds.
The crystals were analyzed using SEM with EDX, Laue backscattering, powder X-ray diffraction and SQUID magnetometry measurements.

\textbf{Elasto-thermoelectric transpot measurements}

We applied an in-plane uniaxial strain by gluing the samples on commercial piezoelectric stacks (Part. No. Pst 150/5x5x7, form Piezomechanik, Munich, Germany), using the Devcon General Purpose Adhesive Epoxy (No. 14250). The samples are oriented in order to align their orthorhombic axis to the strain axis of the piezo, in analogy to elastoresistivity measurements \cite{Chu710,Kuo958}. We built a thermal circuit by connecting through silver wires one side of the sample to a resistive heater and the other side to the thermal mass of our sample holder. Longitudinal and transverse couples of electrodes have been attached to the sample to pick up the Seebeck and the Nernst signals (Figure 1a).  Figure 1b and Figure 1e show the measurement procedure for collecting the strain derivative of the Seebeck and the Nernst anisotropies. Basically, a heating power W$_H$ is applied to the sample (bottom panels in Figure 1c and 1e) in order to create a temperature gradient $\nabla T$=$\Delta T$/$l_T$, where $\Delta T$ is the temperature difference measured through a chromel-Au thermocouple and $l_T$ is the distance between the thermocouple tips. This causes a response in the Seebeck and Nernst voltages ($\Delta V_S$ and $\Delta V_N$ in the top panels of Figure 1b and 1e). The corresponding Seebeck and Nernst coefficients are $S$=-$\Delta V_S$/($l_S$$\nabla T$) and $N$=$\Delta V_N$/($l_N$$\nabla T$), where $l_S$ and $l_N$ are the distance between the respective couples of electrodes. Then, the voltage applied to the piezo is varied along the cycle 0 V $\rightarrow$100 V $\rightarrow$ -30 V $\rightarrow$ 0 V and the effective strain is measured through a strain-gauge mounted to the back-side of the piezo stack (middle panels in Figure  1b and 3e). If the sample is thin enough (thickness $<$60-80 $\mu$m), it has been verified \cite{Chu710} that the strain is fully transmitted in a wide temperature range. Since our samples are typically 20-40 $\mu$m thick, mounting the strain-gauge on the back-side of the piezo represents a reliable method to obtain an accurate measurement of the applied strain. 

\newpage

\bibliography{BibElasto}

\begin{thebibliography}{37}%
\makeatletter
\providecommand \@ifxundefined [1]{%
 \@ifx{#1\undefined}
}%
\providecommand \@ifnum [1]{%
 \ifnum #1\expandafter \@firstoftwo
 \else \expandafter \@secondoftwo
 \fi
}%
\providecommand \@ifx [1]{%
 \ifx #1\expandafter \@firstoftwo
 \else \expandafter \@secondoftwo
 \fi
}%
\providecommand \natexlab [1]{#1}%
\providecommand \enquote  [1]{``#1''}%
\providecommand \bibnamefont  [1]{#1}%
\providecommand \bibfnamefont [1]{#1}%
\providecommand \citenamefont [1]{#1}%
\providecommand \href@noop [0]{\@secondoftwo}%
\providecommand \href [0]{\begingroup \@sanitize@url \@href}%
\providecommand \@href[1]{\@@startlink{#1}\@@href}%
\providecommand \@@href[1]{\endgroup#1\@@endlink}%
\providecommand \@sanitize@url [0]{\catcode `\\12\catcode `\$12\catcode
  `\&12\catcode `\#12\catcode `\^12\catcode `\_12\catcode `\%12\relax}%
\providecommand \@@startlink[1]{}%
\providecommand \@@endlink[0]{}%
\providecommand \url  [0]{\begingroup\@sanitize@url \@url }%
\providecommand \@url [1]{\endgroup\@href {#1}{\urlprefix }}%
\providecommand \urlprefix  [0]{URL }%
\providecommand \Eprint [0]{\href }%
\providecommand \doibase [0]{http://dx.doi.org/}%
\providecommand \selectlanguage [0]{\@gobble}%
\providecommand \bibinfo  [0]{\@secondoftwo}%
\providecommand \bibfield  [0]{\@secondoftwo}%
\providecommand \translation [1]{[#1]}%
\providecommand \BibitemOpen [0]{}%
\providecommand \bibitemStop [0]{}%
\providecommand \bibitemNoStop [0]{.\EOS\space}%
\providecommand \EOS [0]{\spacefactor3000\relax}%
\providecommand \BibitemShut  [1]{\csname bibitem#1\endcsname}%
\let\auto@bib@innerbib\@empty
%</preamble>
\bibitem [{\citenamefont {Daou}\ \emph {et~al.}(2010)\citenamefont {Daou},
  \citenamefont {Chang}, \citenamefont {LeBoeuf}, \citenamefont
  {Cyr-Choini{\`e}re}, \citenamefont {Lalibert{\'e}}, \citenamefont
  {Doiron-Leyraud}, \citenamefont {Ramshaw}, \citenamefont {Liang},
  \citenamefont {Bonn}, \citenamefont {Hardy},\ and\ \citenamefont
  {Taillefer}}]{Daou:2010aa}%
  \BibitemOpen
  \bibfield  {author} {\bibinfo {author} {\bibfnamefont {R.}~\bibnamefont
  {Daou}}, \bibinfo {author} {\bibfnamefont {J.}~\bibnamefont {Chang}},
  \bibinfo {author} {\bibfnamefont {D.}~\bibnamefont {LeBoeuf}}, \bibinfo
  {author} {\bibfnamefont {O.}~\bibnamefont {Cyr-Choini{\`e}re}}, \bibinfo
  {author} {\bibfnamefont {F.}~\bibnamefont {Lalibert{\'e}}}, \bibinfo {author}
  {\bibfnamefont {N.}~\bibnamefont {Doiron-Leyraud}}, \bibinfo {author}
  {\bibfnamefont {B.~J.}\ \bibnamefont {Ramshaw}}, \bibinfo {author}
  {\bibfnamefont {R.}~\bibnamefont {Liang}}, \bibinfo {author} {\bibfnamefont
  {D.~A.}\ \bibnamefont {Bonn}}, \bibinfo {author} {\bibfnamefont {W.~N.}\
  \bibnamefont {Hardy}}, \ and\ \bibinfo {author} {\bibfnamefont
  {L.}~\bibnamefont {Taillefer}},\ }\href {https://doi.org/10.1038/nature08716}
  {\bibfield  {journal} {\bibinfo  {journal} {Nature}\ }\textbf {\bibinfo
  {volume} {463}},\ \bibinfo {pages} {519 EP } (\bibinfo {year}
  {2010})}\BibitemShut {NoStop}%
\bibitem [{\citenamefont {Lawler}\ \emph {et~al.}(2010)\citenamefont {Lawler},
  \citenamefont {Fujita}, \citenamefont {Lee}, \citenamefont {Schmidt},
  \citenamefont {Kohsaka}, \citenamefont {Kim}, \citenamefont {Eisaki},
  \citenamefont {Uchida}, \citenamefont {Davis}, \citenamefont {Sethna},\ and\
  \citenamefont {Kim}}]{Lawler:2010aa}%
  \BibitemOpen
  \bibfield  {author} {\bibinfo {author} {\bibfnamefont {M.~J.}\ \bibnamefont
  {Lawler}}, \bibinfo {author} {\bibfnamefont {K.}~\bibnamefont {Fujita}},
  \bibinfo {author} {\bibfnamefont {J.}~\bibnamefont {Lee}}, \bibinfo {author}
  {\bibfnamefont {A.~R.}\ \bibnamefont {Schmidt}}, \bibinfo {author}
  {\bibfnamefont {Y.}~\bibnamefont {Kohsaka}}, \bibinfo {author} {\bibfnamefont
  {C.~K.}\ \bibnamefont {Kim}}, \bibinfo {author} {\bibfnamefont
  {H.}~\bibnamefont {Eisaki}}, \bibinfo {author} {\bibfnamefont
  {S.}~\bibnamefont {Uchida}}, \bibinfo {author} {\bibfnamefont {J.~C.}\
  \bibnamefont {Davis}}, \bibinfo {author} {\bibfnamefont {J.~P.}\ \bibnamefont
  {Sethna}}, \ and\ \bibinfo {author} {\bibfnamefont {E.-A.}\ \bibnamefont
  {Kim}},\ }\href {https://doi.org/10.1038/nature09169} {\bibfield  {journal}
  {\bibinfo  {journal} {Nature}\ }\textbf {\bibinfo {volume} {466}},\ \bibinfo
  {pages} {347 EP } (\bibinfo {year} {2010})}\BibitemShut {NoStop}%
\bibitem [{\citenamefont {Fernandes}\ \emph {et~al.}(2014)\citenamefont
  {Fernandes}, \citenamefont {Chubukov},\ and\ \citenamefont
  {Schmalian}}]{Fernandes:2014aa}%
  \BibitemOpen
  \bibfield  {author} {\bibinfo {author} {\bibfnamefont {R.~M.}\ \bibnamefont
  {Fernandes}}, \bibinfo {author} {\bibfnamefont {A.~V.}\ \bibnamefont
  {Chubukov}}, \ and\ \bibinfo {author} {\bibfnamefont {J.}~\bibnamefont
  {Schmalian}},\ }\href {https://doi.org/10.1038/nphys2877} {\bibfield
  {journal} {\bibinfo  {journal} {Nature Physics}\ }\textbf {\bibinfo {volume}
  {10}},\ \bibinfo {pages} {97 EP } (\bibinfo {year} {2014})}\BibitemShut
  {NoStop}%
\bibitem [{\citenamefont {Chuang}\ \emph {et~al.}(2010)\citenamefont {Chuang},
  \citenamefont {Allan}, \citenamefont {Lee}, \citenamefont {Xie},
  \citenamefont {Ni}, \citenamefont {Bud{\textquoteright}ko}, \citenamefont
  {Boebinger}, \citenamefont {Canfield},\ and\ \citenamefont
  {Davis}}]{Chuang181}%
  \BibitemOpen
  \bibfield  {author} {\bibinfo {author} {\bibfnamefont {T.-M.}\ \bibnamefont
  {Chuang}}, \bibinfo {author} {\bibfnamefont {M.~P.}\ \bibnamefont {Allan}},
  \bibinfo {author} {\bibfnamefont {J.}~\bibnamefont {Lee}}, \bibinfo {author}
  {\bibfnamefont {Y.}~\bibnamefont {Xie}}, \bibinfo {author} {\bibfnamefont
  {N.}~\bibnamefont {Ni}}, \bibinfo {author} {\bibfnamefont {S.~L.}\
  \bibnamefont {Bud{\textquoteright}ko}}, \bibinfo {author} {\bibfnamefont
  {G.~S.}\ \bibnamefont {Boebinger}}, \bibinfo {author} {\bibfnamefont {P.~C.}\
  \bibnamefont {Canfield}}, \ and\ \bibinfo {author} {\bibfnamefont {J.~C.}\
  \bibnamefont {Davis}},\ }\href {\doibase 10.1126/science.1181083} {\bibfield
  {journal} {\bibinfo  {journal} {Science}\ }\textbf {\bibinfo {volume}
  {327}},\ \bibinfo {pages} {181} (\bibinfo {year} {2010})},\ \Eprint
  {http://arxiv.org/abs/http://science.sciencemag.org/content/327/5962/181.full.pdf}
  {http://science.sciencemag.org/content/327/5962/181.full.pdf} \BibitemShut
  {NoStop}%
\bibitem [{\citenamefont {Chu}\ \emph {et~al.}(2012)\citenamefont {Chu},
  \citenamefont {Kuo}, \citenamefont {Analytis},\ and\ \citenamefont
  {Fisher}}]{Chu710}%
  \BibitemOpen
  \bibfield  {author} {\bibinfo {author} {\bibfnamefont {J.-H.}\ \bibnamefont
  {Chu}}, \bibinfo {author} {\bibfnamefont {H.-H.}\ \bibnamefont {Kuo}},
  \bibinfo {author} {\bibfnamefont {J.~G.}\ \bibnamefont {Analytis}}, \ and\
  \bibinfo {author} {\bibfnamefont {I.~R.}\ \bibnamefont {Fisher}},\ }\href
  {\doibase 10.1126/science.1221713} {\bibfield  {journal} {\bibinfo  {journal}
  {Science}\ }\textbf {\bibinfo {volume} {337}},\ \bibinfo {pages} {710}
  (\bibinfo {year} {2012})},\ \Eprint
  {http://arxiv.org/abs/http://science.sciencemag.org/content/337/6095/710.full.pdf}
  {http://science.sciencemag.org/content/337/6095/710.full.pdf} \BibitemShut
  {NoStop}%
\bibitem [{\citenamefont {Kuo}\ \emph {et~al.}(2016)\citenamefont {Kuo},
  \citenamefont {Chu}, \citenamefont {Palmstrom}, \citenamefont {Kivelson},\
  and\ \citenamefont {Fisher}}]{Kuo958}%
  \BibitemOpen
  \bibfield  {author} {\bibinfo {author} {\bibfnamefont {H.-H.}\ \bibnamefont
  {Kuo}}, \bibinfo {author} {\bibfnamefont {J.-H.}\ \bibnamefont {Chu}},
  \bibinfo {author} {\bibfnamefont {J.~C.}\ \bibnamefont {Palmstrom}}, \bibinfo
  {author} {\bibfnamefont {S.~A.}\ \bibnamefont {Kivelson}}, \ and\ \bibinfo
  {author} {\bibfnamefont {I.~R.}\ \bibnamefont {Fisher}},\ }\href {\doibase
  10.1126/science.aab0103} {\bibfield  {journal} {\bibinfo  {journal}
  {Science}\ }\textbf {\bibinfo {volume} {352}},\ \bibinfo {pages} {958}
  (\bibinfo {year} {2016})},\ \Eprint
  {http://arxiv.org/abs/http://science.sciencemag.org/content/352/6288/958.full.pdf}
  {http://science.sciencemag.org/content/352/6288/958.full.pdf} \BibitemShut
  {NoStop}%
\bibitem [{\citenamefont {Hosoi}\ \emph {et~al.}(2016)\citenamefont {Hosoi},
  \citenamefont {Matsuura}, \citenamefont {Ishida}, \citenamefont {Wang},
  \citenamefont {Mizukami}, \citenamefont {Watashige}, \citenamefont
  {Kasahara}, \citenamefont {Matsuda},\ and\ \citenamefont
  {Shibauchi}}]{Hosoi8139}%
  \BibitemOpen
  \bibfield  {author} {\bibinfo {author} {\bibfnamefont {S.}~\bibnamefont
  {Hosoi}}, \bibinfo {author} {\bibfnamefont {K.}~\bibnamefont {Matsuura}},
  \bibinfo {author} {\bibfnamefont {K.}~\bibnamefont {Ishida}}, \bibinfo
  {author} {\bibfnamefont {H.}~\bibnamefont {Wang}}, \bibinfo {author}
  {\bibfnamefont {Y.}~\bibnamefont {Mizukami}}, \bibinfo {author}
  {\bibfnamefont {T.}~\bibnamefont {Watashige}}, \bibinfo {author}
  {\bibfnamefont {S.}~\bibnamefont {Kasahara}}, \bibinfo {author}
  {\bibfnamefont {Y.}~\bibnamefont {Matsuda}}, \ and\ \bibinfo {author}
  {\bibfnamefont {T.}~\bibnamefont {Shibauchi}},\ }\href {\doibase
  10.1073/pnas.1605806113} {\bibfield  {journal} {\bibinfo  {journal}
  {Proceedings of the National Academy of Sciences}\ }\textbf {\bibinfo
  {volume} {113}},\ \bibinfo {pages} {8139} (\bibinfo {year} {2016})},\ \Eprint
  {http://arxiv.org/abs/https://www.pnas.org/content/113/29/8139.full.pdf}
  {https://www.pnas.org/content/113/29/8139.full.pdf} \BibitemShut {NoStop}%
\bibitem [{\citenamefont {Behnia}(2009)}]{0953-8984-21-11-113101}%
  \BibitemOpen
  \bibfield  {author} {\bibinfo {author} {\bibfnamefont {K.}~\bibnamefont
  {Behnia}},\ }\href {http://stacks.iop.org/0953-8984/21/i=11/a=113101}
  {\bibfield  {journal} {\bibinfo  {journal} {Journal of Physics: Condensed
  Matter}\ }\textbf {\bibinfo {volume} {21}},\ \bibinfo {pages} {113101}
  (\bibinfo {year} {2009})}\BibitemShut {NoStop}%
\bibitem [{\citenamefont {Pallecchi}\ \emph {et~al.}(2016)\citenamefont
  {Pallecchi}, \citenamefont {Caglieris},\ and\ \citenamefont
  {Putti}}]{0953-2048-29-7-073002}%
  \BibitemOpen
  \bibfield  {author} {\bibinfo {author} {\bibfnamefont {I.}~\bibnamefont
  {Pallecchi}}, \bibinfo {author} {\bibfnamefont {F.}~\bibnamefont
  {Caglieris}}, \ and\ \bibinfo {author} {\bibfnamefont {M.}~\bibnamefont
  {Putti}},\ }\href {http://stacks.iop.org/0953-2048/29/i=7/a=073002}
  {\bibfield  {journal} {\bibinfo  {journal} {Superconductor Science and
  Technology}\ }\textbf {\bibinfo {volume} {29}},\ \bibinfo {pages} {073002}
  (\bibinfo {year} {2016})}\BibitemShut {NoStop}%
\bibitem [{\citenamefont {Matusiak}\ \emph {et~al.}(2011)\citenamefont
  {Matusiak}, \citenamefont {Bukowski},\ and\ \citenamefont
  {Karpinski}}]{PhysRevB.83.224505}%
  \BibitemOpen
  \bibfield  {author} {\bibinfo {author} {\bibfnamefont {M.}~\bibnamefont
  {Matusiak}}, \bibinfo {author} {\bibfnamefont {Z.}~\bibnamefont {Bukowski}},
  \ and\ \bibinfo {author} {\bibfnamefont {J.}~\bibnamefont {Karpinski}},\
  }\href {\doibase 10.1103/PhysRevB.83.224505} {\bibfield  {journal} {\bibinfo
  {journal} {Phys. Rev. B}\ }\textbf {\bibinfo {volume} {83}},\ \bibinfo
  {pages} {224505} (\bibinfo {year} {2011})}\BibitemShut {NoStop}%
\bibitem [{\citenamefont {Kondrat}\ \emph {et~al.}(2011)\citenamefont
  {Kondrat}, \citenamefont {Behr}, \citenamefont {B\"uchner},\ and\
  \citenamefont {Hess}}]{PhysRevB.83.092507}%
  \BibitemOpen
  \bibfield  {author} {\bibinfo {author} {\bibfnamefont {A.}~\bibnamefont
  {Kondrat}}, \bibinfo {author} {\bibfnamefont {G.}~\bibnamefont {Behr}},
  \bibinfo {author} {\bibfnamefont {B.}~\bibnamefont {B\"uchner}}, \ and\
  \bibinfo {author} {\bibfnamefont {C.}~\bibnamefont {Hess}},\ }\href {\doibase
  10.1103/PhysRevB.83.092507} {\bibfield  {journal} {\bibinfo  {journal} {Phys.
  Rev. B}\ }\textbf {\bibinfo {volume} {83}},\ \bibinfo {pages} {092507}
  (\bibinfo {year} {2011})}\BibitemShut {NoStop}%
\bibitem [{\citenamefont {Behnia}\ and\ \citenamefont
  {Aubin}(2016)}]{0034-4885-79-4-046502}%
  \BibitemOpen
  \bibfield  {author} {\bibinfo {author} {\bibfnamefont {K.}~\bibnamefont
  {Behnia}}\ and\ \bibinfo {author} {\bibfnamefont {H.}~\bibnamefont {Aubin}},\
  }\href {http://stacks.iop.org/0034-4885/79/i=4/a=046502} {\bibfield
  {journal} {\bibinfo  {journal} {Reports on Progress in Physics}\ }\textbf
  {\bibinfo {volume} {79}},\ \bibinfo {pages} {046502} (\bibinfo {year}
  {2016})}\BibitemShut {NoStop}%
\bibitem [{\citenamefont {Fujii}\ \emph {et~al.}(2016)\citenamefont {Fujii},
  \citenamefont {Shirachi},\ and\ \citenamefont {Asamitsu}}]{FUJII201631}%
  \BibitemOpen
  \bibfield  {author} {\bibinfo {author} {\bibfnamefont {T.}~\bibnamefont
  {Fujii}}, \bibinfo {author} {\bibfnamefont {T.}~\bibnamefont {Shirachi}}, \
  and\ \bibinfo {author} {\bibfnamefont {A.}~\bibnamefont {Asamitsu}},\ }\href
  {\doibase https://doi.org/10.1016/j.physc.2016.03.007} {\bibfield  {journal}
  {\bibinfo  {journal} {Physica C: Superconductivity and its Applications}\
  }\textbf {\bibinfo {volume} {530}},\ \bibinfo {pages} {31 } (\bibinfo {year}
  {2016})},\ \bibinfo {note} {28th International Symposium on
  Superconductivity}\BibitemShut {NoStop}%
\bibitem [{\citenamefont {Jiang}\ \emph {et~al.}(2013)\citenamefont {Jiang},
  \citenamefont {Jeevan}, \citenamefont {Dong},\ and\ \citenamefont
  {Gegenwart}}]{PhysRevLett.110.067001}%
  \BibitemOpen
  \bibfield  {author} {\bibinfo {author} {\bibfnamefont {S.}~\bibnamefont
  {Jiang}}, \bibinfo {author} {\bibfnamefont {H.~S.}\ \bibnamefont {Jeevan}},
  \bibinfo {author} {\bibfnamefont {J.}~\bibnamefont {Dong}}, \ and\ \bibinfo
  {author} {\bibfnamefont {P.}~\bibnamefont {Gegenwart}},\ }\href {\doibase
  10.1103/PhysRevLett.110.067001} {\bibfield  {journal} {\bibinfo  {journal}
  {Phys. Rev. Lett.}\ }\textbf {\bibinfo {volume} {110}},\ \bibinfo {pages}
  {067001} (\bibinfo {year} {2013})}\BibitemShut {NoStop}%
\bibitem [{\citenamefont {Matusiak}\ \emph
  {et~al.}(2018{\natexlab{a}})\citenamefont {Matusiak}, \citenamefont {Babij},\
  and\ \citenamefont {Wolf}}]{PhysRevB.97.100506}%
  \BibitemOpen
  \bibfield  {author} {\bibinfo {author} {\bibfnamefont {M.}~\bibnamefont
  {Matusiak}}, \bibinfo {author} {\bibfnamefont {M.}~\bibnamefont {Babij}}, \
  and\ \bibinfo {author} {\bibfnamefont {T.}~\bibnamefont {Wolf}},\ }\href
  {\doibase 10.1103/PhysRevB.97.100506} {\bibfield  {journal} {\bibinfo
  {journal} {Phys. Rev. B}\ }\textbf {\bibinfo {volume} {97}},\ \bibinfo
  {pages} {100506} (\bibinfo {year} {2018}{\natexlab{a}})}\BibitemShut
  {NoStop}%
\bibitem [{\citenamefont {Matusiak}\ \emph
  {et~al.}(2018{\natexlab{b}})\citenamefont {Matusiak}, \citenamefont
  {Rogacki},\ and\ \citenamefont {Wolf}}]{PhysRevB.97.220501}%
  \BibitemOpen
  \bibfield  {author} {\bibinfo {author} {\bibfnamefont {M.}~\bibnamefont
  {Matusiak}}, \bibinfo {author} {\bibfnamefont {K.}~\bibnamefont {Rogacki}}, \
  and\ \bibinfo {author} {\bibfnamefont {T.}~\bibnamefont {Wolf}},\ }\href
  {\doibase 10.1103/PhysRevB.97.220501} {\bibfield  {journal} {\bibinfo
  {journal} {Phys. Rev. B}\ }\textbf {\bibinfo {volume} {97}},\ \bibinfo
  {pages} {220501} (\bibinfo {year} {2018}{\natexlab{b}})}\BibitemShut
  {NoStop}%
\bibitem [{\citenamefont {Kappenberger}\ \emph {et~al.}(2018)\citenamefont
  {Kappenberger}, \citenamefont {Aswartham}, \citenamefont {Scaravaggi},
  \citenamefont {Blum}, \citenamefont {Sturza}, \citenamefont {Wolter},
  \citenamefont {Wurmehl},\ and\ \citenamefont
  {B{\"u}chner}}]{KAPPENBERGER20189}%
  \BibitemOpen
  \bibfield  {author} {\bibinfo {author} {\bibfnamefont {R.}~\bibnamefont
  {Kappenberger}}, \bibinfo {author} {\bibfnamefont {S.}~\bibnamefont
  {Aswartham}}, \bibinfo {author} {\bibfnamefont {F.}~\bibnamefont
  {Scaravaggi}}, \bibinfo {author} {\bibfnamefont {C.~G.}\ \bibnamefont
  {Blum}}, \bibinfo {author} {\bibfnamefont {M.~I.}\ \bibnamefont {Sturza}},
  \bibinfo {author} {\bibfnamefont {A.~U.}\ \bibnamefont {Wolter}}, \bibinfo
  {author} {\bibfnamefont {S.}~\bibnamefont {Wurmehl}}, \ and\ \bibinfo
  {author} {\bibfnamefont {B.}~\bibnamefont {B{\"u}chner}},\ }\href {\doibase
  https://doi.org/10.1016/j.jcrysgro.2017.11.006} {\bibfield  {journal}
  {\bibinfo  {journal} {Journal of Crystal Growth}\ }\textbf {\bibinfo {volume}
  {483}},\ \bibinfo {pages} {9 } (\bibinfo {year} {2018})}\BibitemShut
  {NoStop}%
\bibitem [{\citenamefont {Wang}\ \emph {et~al.}(2019)\citenamefont {Wang},
  \citenamefont {Sauerland}, \citenamefont {Scaravaggi}, \citenamefont
  {Kappenberger}, \citenamefont {Aswartham}, \citenamefont {Wurmehl},
  \citenamefont {Wolter}, \citenamefont {B{\"u}chner},\ and\ \citenamefont
  {Klingeler}}]{WANG201950}%
  \BibitemOpen
  \bibfield  {author} {\bibinfo {author} {\bibfnamefont {L.}~\bibnamefont
  {Wang}}, \bibinfo {author} {\bibfnamefont {S.}~\bibnamefont {Sauerland}},
  \bibinfo {author} {\bibfnamefont {F.}~\bibnamefont {Scaravaggi}}, \bibinfo
  {author} {\bibfnamefont {R.}~\bibnamefont {Kappenberger}}, \bibinfo {author}
  {\bibfnamefont {S.}~\bibnamefont {Aswartham}}, \bibinfo {author}
  {\bibfnamefont {S.}~\bibnamefont {Wurmehl}}, \bibinfo {author} {\bibfnamefont
  {A.}~\bibnamefont {Wolter}}, \bibinfo {author} {\bibfnamefont
  {B.}~\bibnamefont {B{\"u}chner}}, \ and\ \bibinfo {author} {\bibfnamefont
  {R.}~\bibnamefont {Klingeler}},\ }\href {\doibase
  https://doi.org/10.1016/j.jmmm.2019.02.061} {\bibfield  {journal} {\bibinfo
  {journal} {Journal of Magnetism and Magnetic Materials}\ }\textbf {\bibinfo
  {volume} {482}},\ \bibinfo {pages} {50 } (\bibinfo {year}
  {2019})}\BibitemShut {NoStop}%
\bibitem [{\citenamefont {Hackl}\ and\ \citenamefont
  {Vojta}(2009)}]{PhysRevB.80.220514}%
  \BibitemOpen
  \bibfield  {author} {\bibinfo {author} {\bibfnamefont {A.}~\bibnamefont
  {Hackl}}\ and\ \bibinfo {author} {\bibfnamefont {M.}~\bibnamefont {Vojta}},\
  }\href {\doibase 10.1103/PhysRevB.80.220514} {\bibfield  {journal} {\bibinfo
  {journal} {Phys. Rev. B}\ }\textbf {\bibinfo {volume} {80}},\ \bibinfo
  {pages} {220514} (\bibinfo {year} {2009})}\BibitemShut {NoStop}%
\bibitem [{\citenamefont {Wang}\ \emph {et~al.}(2009)\citenamefont {Wang},
  \citenamefont {Li}, \citenamefont {Zhu}, \citenamefont {Jiang}, \citenamefont
  {Lin}, \citenamefont {Luo}, \citenamefont {Chi}, \citenamefont {Li},
  \citenamefont {Ren}, \citenamefont {He}, \citenamefont {Chen}, \citenamefont
  {Wang}, \citenamefont {Tao}, \citenamefont {Cao},\ and\ \citenamefont
  {Xu}}]{PhysRevB.79.054521}%
  \BibitemOpen
  \bibfield  {author} {\bibinfo {author} {\bibfnamefont {C.}~\bibnamefont
  {Wang}}, \bibinfo {author} {\bibfnamefont {Y.~K.}\ \bibnamefont {Li}},
  \bibinfo {author} {\bibfnamefont {Z.~W.}\ \bibnamefont {Zhu}}, \bibinfo
  {author} {\bibfnamefont {S.}~\bibnamefont {Jiang}}, \bibinfo {author}
  {\bibfnamefont {X.}~\bibnamefont {Lin}}, \bibinfo {author} {\bibfnamefont
  {Y.~K.}\ \bibnamefont {Luo}}, \bibinfo {author} {\bibfnamefont
  {S.}~\bibnamefont {Chi}}, \bibinfo {author} {\bibfnamefont {L.~J.}\
  \bibnamefont {Li}}, \bibinfo {author} {\bibfnamefont {Z.}~\bibnamefont
  {Ren}}, \bibinfo {author} {\bibfnamefont {M.}~\bibnamefont {He}}, \bibinfo
  {author} {\bibfnamefont {H.}~\bibnamefont {Chen}}, \bibinfo {author}
  {\bibfnamefont {Y.~T.}\ \bibnamefont {Wang}}, \bibinfo {author}
  {\bibfnamefont {Q.}~\bibnamefont {Tao}}, \bibinfo {author} {\bibfnamefont
  {G.~H.}\ \bibnamefont {Cao}}, \ and\ \bibinfo {author} {\bibfnamefont
  {Z.~A.}\ \bibnamefont {Xu}},\ }\href {\doibase 10.1103/PhysRevB.79.054521}
  {\bibfield  {journal} {\bibinfo  {journal} {Phys. Rev. B}\ }\textbf {\bibinfo
  {volume} {79}},\ \bibinfo {pages} {054521} (\bibinfo {year}
  {2009})}\BibitemShut {NoStop}%
\bibitem [{\citenamefont {Okram}\ \emph {et~al.}(2012)\citenamefont {Okram},
  \citenamefont {Kaurav}, \citenamefont {Soni}, \citenamefont {Pal},\ and\
  \citenamefont {Awana}}]{doi:10.1063/1.4766936}%
  \BibitemOpen
  \bibfield  {author} {\bibinfo {author} {\bibfnamefont {G.~S.}\ \bibnamefont
  {Okram}}, \bibinfo {author} {\bibfnamefont {N.}~\bibnamefont {Kaurav}},
  \bibinfo {author} {\bibfnamefont {A.}~\bibnamefont {Soni}}, \bibinfo {author}
  {\bibfnamefont {A.}~\bibnamefont {Pal}}, \ and\ \bibinfo {author}
  {\bibfnamefont {V.~P.~S.}\ \bibnamefont {Awana}},\ }\href {\doibase
  10.1063/1.4766936} {\bibfield  {journal} {\bibinfo  {journal} {AIP Advances}\
  }\textbf {\bibinfo {volume} {2}},\ \bibinfo {pages} {042137} (\bibinfo {year}
  {2012})},\ \Eprint {http://arxiv.org/abs/https://doi.org/10.1063/1.4766936}
  {https://doi.org/10.1063/1.4766936} \BibitemShut {NoStop}%
\bibitem [{\citenamefont {Hong}\ \emph {et~al.}(2020)\citenamefont {Hong},
  \citenamefont {Caglieris}, \citenamefont {Kappenberger}, \citenamefont
  {Wurmehl}, \citenamefont {Aswartham}, \citenamefont {Scaravaggi},
  \citenamefont {Lepucki}, \citenamefont {Wolter}, \citenamefont {Grafe},
  \citenamefont {B\"uchner},\ and\ \citenamefont
  {Hess}}]{PhysRevLett.125.067001}%
  \BibitemOpen
  \bibfield  {author} {\bibinfo {author} {\bibfnamefont {X.}~\bibnamefont
  {Hong}}, \bibinfo {author} {\bibfnamefont {F.}~\bibnamefont {Caglieris}},
  \bibinfo {author} {\bibfnamefont {R.}~\bibnamefont {Kappenberger}}, \bibinfo
  {author} {\bibfnamefont {S.}~\bibnamefont {Wurmehl}}, \bibinfo {author}
  {\bibfnamefont {S.}~\bibnamefont {Aswartham}}, \bibinfo {author}
  {\bibfnamefont {F.}~\bibnamefont {Scaravaggi}}, \bibinfo {author}
  {\bibfnamefont {P.}~\bibnamefont {Lepucki}}, \bibinfo {author} {\bibfnamefont
  {A.~U.~B.}\ \bibnamefont {Wolter}}, \bibinfo {author} {\bibfnamefont {H.-J.}\
  \bibnamefont {Grafe}}, \bibinfo {author} {\bibfnamefont {B.}~\bibnamefont
  {B\"uchner}}, \ and\ \bibinfo {author} {\bibfnamefont {C.}~\bibnamefont
  {Hess}},\ }\href {\doibase 10.1103/PhysRevLett.125.067001} {\bibfield
  {journal} {\bibinfo  {journal} {Phys. Rev. Lett.}\ }\textbf {\bibinfo
  {volume} {125}},\ \bibinfo {pages} {067001} (\bibinfo {year}
  {2020})}\BibitemShut {NoStop}%
\bibitem [{\citenamefont {Bel}\ \emph {et~al.}(2003)\citenamefont {Bel},
  \citenamefont {Behnia},\ and\ \citenamefont
  {Berger}}]{PhysRevLett.91.066602}%
  \BibitemOpen
  \bibfield  {author} {\bibinfo {author} {\bibfnamefont {R.}~\bibnamefont
  {Bel}}, \bibinfo {author} {\bibfnamefont {K.}~\bibnamefont {Behnia}}, \ and\
  \bibinfo {author} {\bibfnamefont {H.}~\bibnamefont {Berger}},\ }\href
  {\doibase 10.1103/PhysRevLett.91.066602} {\bibfield  {journal} {\bibinfo
  {journal} {Phys. Rev. Lett.}\ }\textbf {\bibinfo {volume} {91}},\ \bibinfo
  {pages} {066602} (\bibinfo {year} {2003})}\BibitemShut {NoStop}%
\bibitem [{\citenamefont {Hong}\ \emph {et~al.}(2019)\citenamefont {Hong},
  \citenamefont {Caglieris}, \citenamefont {Kappenberger}, \citenamefont
  {Wurmehl}, \citenamefont {Aswartham}, \citenamefont {B{\"u}chner},\ and\
  \citenamefont {Hess}}]{XC-arxiv}%
  \BibitemOpen
  \bibfield  {author} {\bibinfo {author} {\bibfnamefont {X.~C.}\ \bibnamefont
  {Hong}}, \bibinfo {author} {\bibfnamefont {F.}~\bibnamefont {Caglieris}},
  \bibinfo {author} {\bibfnamefont {R.}~\bibnamefont {Kappenberger}}, \bibinfo
  {author} {\bibfnamefont {S.}~\bibnamefont {Wurmehl}}, \bibinfo {author}
  {\bibfnamefont {S.}~\bibnamefont {Aswartham}}, \bibinfo {author}
  {\bibfnamefont {B.}~\bibnamefont {B{\"u}chner}}, \ and\ \bibinfo {author}
  {\bibfnamefont {C.}~\bibnamefont {Hess}},\ }\href@noop {} {\bibfield
  {journal} {\bibinfo  {journal} {arXiv:1908.00484v1}\ } (\bibinfo {year}
  {2019})}\BibitemShut {NoStop}%
\bibitem [{\citenamefont {Meinero}\ \emph {et~al.}(2018)\citenamefont
  {Meinero}, \citenamefont {Caglieris}, \citenamefont {Lamura}, \citenamefont
  {Pallecchi}, \citenamefont {Jost}, \citenamefont {Zeitler}, \citenamefont
  {Ishida}, \citenamefont {Eisaki},\ and\ \citenamefont
  {Putti}}]{PhysRevB.98.155116}%
  \BibitemOpen
  \bibfield  {author} {\bibinfo {author} {\bibfnamefont {M.}~\bibnamefont
  {Meinero}}, \bibinfo {author} {\bibfnamefont {F.}~\bibnamefont {Caglieris}},
  \bibinfo {author} {\bibfnamefont {G.}~\bibnamefont {Lamura}}, \bibinfo
  {author} {\bibfnamefont {I.}~\bibnamefont {Pallecchi}}, \bibinfo {author}
  {\bibfnamefont {A.}~\bibnamefont {Jost}}, \bibinfo {author} {\bibfnamefont
  {U.}~\bibnamefont {Zeitler}}, \bibinfo {author} {\bibfnamefont
  {S.}~\bibnamefont {Ishida}}, \bibinfo {author} {\bibfnamefont
  {H.}~\bibnamefont {Eisaki}}, \ and\ \bibinfo {author} {\bibfnamefont
  {M.}~\bibnamefont {Putti}},\ }\href {\doibase 10.1103/PhysRevB.98.155116}
  {\bibfield  {journal} {\bibinfo  {journal} {Phys. Rev. B}\ }\textbf {\bibinfo
  {volume} {98}},\ \bibinfo {pages} {155116} (\bibinfo {year}
  {2018})}\BibitemShut {NoStop}%
\bibitem [{\citenamefont {Fernandes}\ \emph {et~al.}(2011)\citenamefont
  {Fernandes}, \citenamefont {Abrahams},\ and\ \citenamefont
  {Schmalian}}]{PhysRevLett.107.217002}%
  \BibitemOpen
  \bibfield  {author} {\bibinfo {author} {\bibfnamefont {R.~M.}\ \bibnamefont
  {Fernandes}}, \bibinfo {author} {\bibfnamefont {E.}~\bibnamefont {Abrahams}},
  \ and\ \bibinfo {author} {\bibfnamefont {J.}~\bibnamefont {Schmalian}},\
  }\href {\doibase 10.1103/PhysRevLett.107.217002} {\bibfield  {journal}
  {\bibinfo  {journal} {Phys. Rev. Lett.}\ }\textbf {\bibinfo {volume} {107}},\
  \bibinfo {pages} {217002} (\bibinfo {year} {2011})}\BibitemShut {NoStop}%
\bibitem [{\citenamefont {Fernandes}\ \emph {et~al.}(2010)\citenamefont
  {Fernandes}, \citenamefont {VanBebber}, \citenamefont {Bhattacharya},
  \citenamefont {Chandra}, \citenamefont {Keppens}, \citenamefont {Mandrus},
  \citenamefont {McGuire}, \citenamefont {Sales}, \citenamefont {Sefat},\ and\
  \citenamefont {Schmalian}}]{PhysRevLett.105.157003}%
  \BibitemOpen
  \bibfield  {author} {\bibinfo {author} {\bibfnamefont {R.~M.}\ \bibnamefont
  {Fernandes}}, \bibinfo {author} {\bibfnamefont {L.~H.}\ \bibnamefont
  {VanBebber}}, \bibinfo {author} {\bibfnamefont {S.}~\bibnamefont
  {Bhattacharya}}, \bibinfo {author} {\bibfnamefont {P.}~\bibnamefont
  {Chandra}}, \bibinfo {author} {\bibfnamefont {V.}~\bibnamefont {Keppens}},
  \bibinfo {author} {\bibfnamefont {D.}~\bibnamefont {Mandrus}}, \bibinfo
  {author} {\bibfnamefont {M.~A.}\ \bibnamefont {McGuire}}, \bibinfo {author}
  {\bibfnamefont {B.~C.}\ \bibnamefont {Sales}}, \bibinfo {author}
  {\bibfnamefont {A.~S.}\ \bibnamefont {Sefat}}, \ and\ \bibinfo {author}
  {\bibfnamefont {J.}~\bibnamefont {Schmalian}},\ }\href {\doibase
  10.1103/PhysRevLett.105.157003} {\bibfield  {journal} {\bibinfo  {journal}
  {Phys. Rev. Lett.}\ }\textbf {\bibinfo {volume} {105}},\ \bibinfo {pages}
  {157003} (\bibinfo {year} {2010})}\BibitemShut {NoStop}%
\bibitem [{\citenamefont {Breitkreiz}\ \emph {et~al.}(2014)\citenamefont
  {Breitkreiz}, \citenamefont {Brydon},\ and\ \citenamefont
  {Timm}}]{PhysRevB.90.121104}%
  \BibitemOpen
  \bibfield  {author} {\bibinfo {author} {\bibfnamefont {M.}~\bibnamefont
  {Breitkreiz}}, \bibinfo {author} {\bibfnamefont {P.~M.~R.}\ \bibnamefont
  {Brydon}}, \ and\ \bibinfo {author} {\bibfnamefont {C.}~\bibnamefont
  {Timm}},\ }\href {\doibase 10.1103/PhysRevB.90.121104} {\bibfield  {journal}
  {\bibinfo  {journal} {Phys. Rev. B}\ }\textbf {\bibinfo {volume} {90}},\
  \bibinfo {pages} {121104} (\bibinfo {year} {2014})}\BibitemShut {NoStop}%
\bibitem [{\citenamefont {Gastiasoro}\ \emph {et~al.}(2014)\citenamefont
  {Gastiasoro}, \citenamefont {Paul}, \citenamefont {Wang}, \citenamefont
  {Hirschfeld},\ and\ \citenamefont {Andersen}}]{PhysRevLett.113.127001}%
  \BibitemOpen
  \bibfield  {author} {\bibinfo {author} {\bibfnamefont {M.~N.}\ \bibnamefont
  {Gastiasoro}}, \bibinfo {author} {\bibfnamefont {I.}~\bibnamefont {Paul}},
  \bibinfo {author} {\bibfnamefont {Y.}~\bibnamefont {Wang}}, \bibinfo {author}
  {\bibfnamefont {P.~J.}\ \bibnamefont {Hirschfeld}}, \ and\ \bibinfo {author}
  {\bibfnamefont {B.~M.}\ \bibnamefont {Andersen}},\ }\href {\doibase
  10.1103/PhysRevLett.113.127001} {\bibfield  {journal} {\bibinfo  {journal}
  {Phys. Rev. Lett.}\ }\textbf {\bibinfo {volume} {113}},\ \bibinfo {pages}
  {127001} (\bibinfo {year} {2014})}\BibitemShut {NoStop}%
\bibitem [{\citenamefont {Kontani}\ \emph {et~al.}(2011)\citenamefont
  {Kontani}, \citenamefont {Saito},\ and\ \citenamefont
  {Onari}}]{PhysRevB.84.024528}%
  \BibitemOpen
  \bibfield  {author} {\bibinfo {author} {\bibfnamefont {H.}~\bibnamefont
  {Kontani}}, \bibinfo {author} {\bibfnamefont {T.}~\bibnamefont {Saito}}, \
  and\ \bibinfo {author} {\bibfnamefont {S.}~\bibnamefont {Onari}},\ }\href
  {\doibase 10.1103/PhysRevB.84.024528} {\bibfield  {journal} {\bibinfo
  {journal} {Phys. Rev. B}\ }\textbf {\bibinfo {volume} {84}},\ \bibinfo
  {pages} {024528} (\bibinfo {year} {2011})}\BibitemShut {NoStop}%
\bibitem [{\citenamefont {Lee}\ \emph {et~al.}(2009)\citenamefont {Lee},
  \citenamefont {Yin},\ and\ \citenamefont {Ku}}]{PhysRevLett.103.267001}%
  \BibitemOpen
  \bibfield  {author} {\bibinfo {author} {\bibfnamefont {C.-C.}\ \bibnamefont
  {Lee}}, \bibinfo {author} {\bibfnamefont {W.-G.}\ \bibnamefont {Yin}}, \ and\
  \bibinfo {author} {\bibfnamefont {W.}~\bibnamefont {Ku}},\ }\href {\doibase
  10.1103/PhysRevLett.103.267001} {\bibfield  {journal} {\bibinfo  {journal}
  {Phys. Rev. Lett.}\ }\textbf {\bibinfo {volume} {103}},\ \bibinfo {pages}
  {267001} (\bibinfo {year} {2009})}\BibitemShut {NoStop}%
\bibitem [{\citenamefont {Valenzuela}\ \emph {et~al.}(2010)\citenamefont
  {Valenzuela}, \citenamefont {Bascones},\ and\ \citenamefont
  {Calder\'on}}]{PhysRevLett.105.207202}%
  \BibitemOpen
  \bibfield  {author} {\bibinfo {author} {\bibfnamefont {B.}~\bibnamefont
  {Valenzuela}}, \bibinfo {author} {\bibfnamefont {E.}~\bibnamefont
  {Bascones}}, \ and\ \bibinfo {author} {\bibfnamefont {M.~J.}\ \bibnamefont
  {Calder\'on}},\ }\href {\doibase 10.1103/PhysRevLett.105.207202} {\bibfield
  {journal} {\bibinfo  {journal} {Phys. Rev. Lett.}\ }\textbf {\bibinfo
  {volume} {105}},\ \bibinfo {pages} {207202} (\bibinfo {year}
  {2010})}\BibitemShut {NoStop}%
\bibitem [{\citenamefont {Chen}\ \emph {et~al.}(2010)\citenamefont {Chen},
  \citenamefont {Maciejko}, \citenamefont {Sorini}, \citenamefont {Moritz},
  \citenamefont {Singh},\ and\ \citenamefont {Devereaux}}]{PhysRevB.82.100504}%
  \BibitemOpen
  \bibfield  {author} {\bibinfo {author} {\bibfnamefont {C.-C.}\ \bibnamefont
  {Chen}}, \bibinfo {author} {\bibfnamefont {J.}~\bibnamefont {Maciejko}},
  \bibinfo {author} {\bibfnamefont {A.~P.}\ \bibnamefont {Sorini}}, \bibinfo
  {author} {\bibfnamefont {B.}~\bibnamefont {Moritz}}, \bibinfo {author}
  {\bibfnamefont {R.~R.~P.}\ \bibnamefont {Singh}}, \ and\ \bibinfo {author}
  {\bibfnamefont {T.~P.}\ \bibnamefont {Devereaux}},\ }\href {\doibase
  10.1103/PhysRevB.82.100504} {\bibfield  {journal} {\bibinfo  {journal} {Phys.
  Rev. B}\ }\textbf {\bibinfo {volume} {82}},\ \bibinfo {pages} {100504}
  (\bibinfo {year} {2010})}\BibitemShut {NoStop}%
\bibitem [{\citenamefont {Ok}\ \emph {et~al.}(2018)\citenamefont {Ok},
  \citenamefont {Baek}, \citenamefont {Efremov}, \citenamefont {Kappenberger},
  \citenamefont {Aswartham}, \citenamefont {Kim}, \citenamefont {van~den
  Brink},\ and\ \citenamefont {B\"uchner}}]{PhysRevB.97.180405}%
  \BibitemOpen
  \bibfield  {author} {\bibinfo {author} {\bibfnamefont {J.~M.}\ \bibnamefont
  {Ok}}, \bibinfo {author} {\bibfnamefont {S.-H.}\ \bibnamefont {Baek}},
  \bibinfo {author} {\bibfnamefont {D.~V.}\ \bibnamefont {Efremov}}, \bibinfo
  {author} {\bibfnamefont {R.}~\bibnamefont {Kappenberger}}, \bibinfo {author}
  {\bibfnamefont {S.}~\bibnamefont {Aswartham}}, \bibinfo {author}
  {\bibfnamefont {J.~S.}\ \bibnamefont {Kim}}, \bibinfo {author} {\bibfnamefont
  {J.}~\bibnamefont {van~den Brink}}, \ and\ \bibinfo {author} {\bibfnamefont
  {B.}~\bibnamefont {B\"uchner}},\ }\href {\doibase 10.1103/PhysRevB.97.180405}
  {\bibfield  {journal} {\bibinfo  {journal} {Phys. Rev. B}\ }\textbf {\bibinfo
  {volume} {97}},\ \bibinfo {pages} {180405} (\bibinfo {year}
  {2018})}\BibitemShut {NoStop}%
\bibitem [{\citenamefont {Fanfarillo}\ \emph {et~al.}(2018)\citenamefont
  {Fanfarillo}, \citenamefont {Benfatto},\ and\ \citenamefont
  {Valenzuela}}]{PhysRevB.97.121109}%
  \BibitemOpen
  \bibfield  {author} {\bibinfo {author} {\bibfnamefont {L.}~\bibnamefont
  {Fanfarillo}}, \bibinfo {author} {\bibfnamefont {L.}~\bibnamefont
  {Benfatto}}, \ and\ \bibinfo {author} {\bibfnamefont {B.}~\bibnamefont
  {Valenzuela}},\ }\href {\doibase 10.1103/PhysRevB.97.121109} {\bibfield
  {journal} {\bibinfo  {journal} {Phys. Rev. B}\ }\textbf {\bibinfo {volume}
  {97}},\ \bibinfo {pages} {121109} (\bibinfo {year} {2018})}\BibitemShut
  {NoStop}%
\bibitem [{\citenamefont {Fern\'andez-Mart\'{\i}n}\ \emph
  {et~al.}(2019)\citenamefont {Fern\'andez-Mart\'{\i}n}, \citenamefont
  {Fanfarillo}, \citenamefont {Benfatto},\ and\ \citenamefont
  {Valenzuela}}]{PhysRevB.99.155117}%
  \BibitemOpen
  \bibfield  {author} {\bibinfo {author} {\bibfnamefont {R.}~\bibnamefont
  {Fern\'andez-Mart\'{\i}n}}, \bibinfo {author} {\bibfnamefont
  {L.}~\bibnamefont {Fanfarillo}}, \bibinfo {author} {\bibfnamefont
  {L.}~\bibnamefont {Benfatto}}, \ and\ \bibinfo {author} {\bibfnamefont
  {B.}~\bibnamefont {Valenzuela}},\ }\href {\doibase
  10.1103/PhysRevB.99.155117} {\bibfield  {journal} {\bibinfo  {journal} {Phys.
  Rev. B}\ }\textbf {\bibinfo {volume} {99}},\ \bibinfo {pages} {155117}
  (\bibinfo {year} {2019})}\BibitemShut {NoStop}%
\bibitem [{\citenamefont {Fanfarillo}\ \emph {et~al.}(2016)\citenamefont
  {Fanfarillo}, \citenamefont {Mansart}, \citenamefont {Toulemonde},
  \citenamefont {Cercellier}, \citenamefont {Le~F\`evre}, \citenamefont
  {Bertran}, \citenamefont {Valenzuela}, \citenamefont {Benfatto},\ and\
  \citenamefont {Brouet}}]{PhysRevB.94.155138}%
  \BibitemOpen
  \bibfield  {author} {\bibinfo {author} {\bibfnamefont {L.}~\bibnamefont
  {Fanfarillo}}, \bibinfo {author} {\bibfnamefont {J.}~\bibnamefont {Mansart}},
  \bibinfo {author} {\bibfnamefont {P.}~\bibnamefont {Toulemonde}}, \bibinfo
  {author} {\bibfnamefont {H.}~\bibnamefont {Cercellier}}, \bibinfo {author}
  {\bibfnamefont {P.}~\bibnamefont {Le~F\`evre}}, \bibinfo {author}
  {\bibfnamefont {F.~m.~c.}\ \bibnamefont {Bertran}}, \bibinfo {author}
  {\bibfnamefont {B.}~\bibnamefont {Valenzuela}}, \bibinfo {author}
  {\bibfnamefont {L.}~\bibnamefont {Benfatto}}, \ and\ \bibinfo {author}
  {\bibfnamefont {V.}~\bibnamefont {Brouet}},\ }\href {\doibase
  10.1103/PhysRevB.94.155138} {\bibfield  {journal} {\bibinfo  {journal} {Phys.
  Rev. B}\ }\textbf {\bibinfo {volume} {94}},\ \bibinfo {pages} {155138}
  (\bibinfo {year} {2016})}\BibitemShut {NoStop}%
\end{thebibliography}%

\newpage

\textbf{Acknowledgments} The authors thank R. Wachtel, D. Meiler, J. Werner, L. Giebeler, S. M\"uller-Litvanyi, and S. Gass (all IFW Dresden) for support. FC thanks Laura Fanfarillo for valuable scientific discussion. This work has been supported by the Deutsche Forschungsgemeinschaft (DFG) through the Priority Programme SPP1458 (Grant No. BU887/15-1), under grant DFG-GRK1621, and through the Emmy Noether Programme WU595/3-3 (S.W.). This project has been supported by the Deutsche Forschungsgemeinschaft through the Research Projects CA 1931/1-1 (F.C.) and AS 523/3-1 (S.A.). This project has received funding from the European Research Council (ERC) under the European Unions' Horizon 2020 research and innovation programme (grant agreement No 647276 - MARS - ERC-2014-CoG).\\

\textbf{Author contributions} \\

\textbf{Competing Interests} The authors declare that they have no competing financial interests.\\

\textbf{Correspondence} Correspondence should be addressed to F. Caglieris~(email: f.caglieris@ifw-dresden.de).\\

\newpage

\begin{figure}[!t]
\includegraphics[width=\columnwidth]{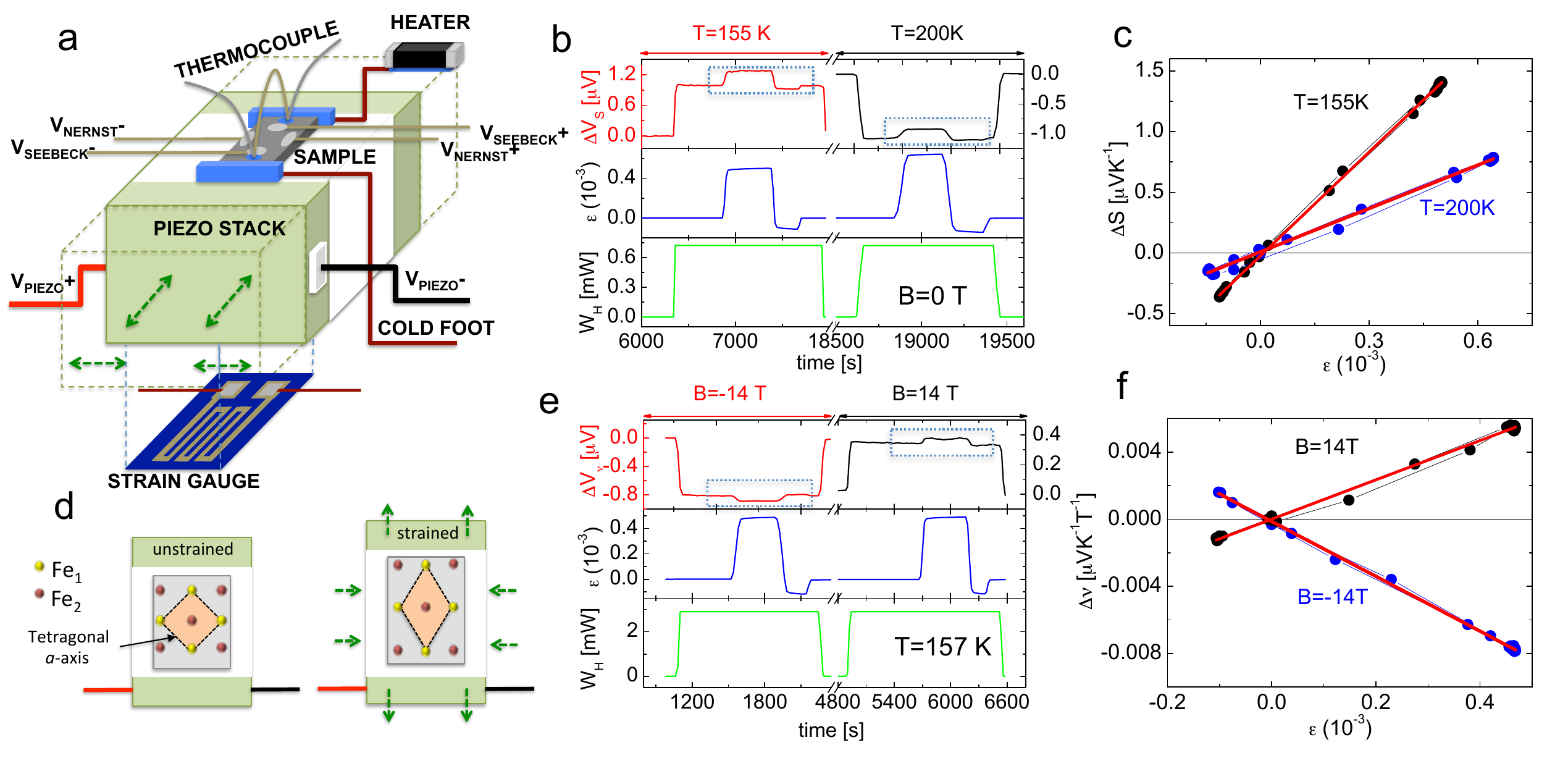}
\caption{\label{Figure} a) Schematic of our experimental setup. The sample is mounted on a piezoelectric device which expands or contracts according to the green arrows. The thermal circuit is realized with a heater and a cold foot connected to the thermal mass. The temperature gradient is measured using a chromel-Au thermocouple while two couples of electrodes measure the Seebeck voltage $\Delta V_S$ and the Nernst voltage $\Delta V_N$. The applied strain $\epsilon$ is measured with a strain-gauge glued on the back side of the piezo. d) The sample is oriented with the tetragonal $a$-axis tilted of 45 degrees with respect to the piezo straining axis, so that the distortion occurs along the putative orthorhombic axis. Fe$_1$ and Fe$_2$ are the iron atoms in the Fe-As planes. b) and e) Time-dependence of the applied heat power $W_H$ (bottom panel), applied strain $\epsilon$ (middle panel) and resulting Seebeck b) and Nernst e) signals at representative temperatures and magnetic fields for a LaFeAsO compound. c) Strain-dependence of the Seebeck anisotropy $\Delta S=S(\epsilon)-S(\epsilon=0)$ for a single crystal of LaFeAsO at $T$=155 K and $T$=200 K. f) Strain-dependence of the Nernst anisotropy $\Delta \nu=\nu(\epsilon)-\nu(\epsilon=0)$ for a single crystal of LaFeAsO at $T$=157 K and $B=\pm14$ T. }
\end{figure}

\begin{figure*}[!t]
\includegraphics[width=\columnwidth]{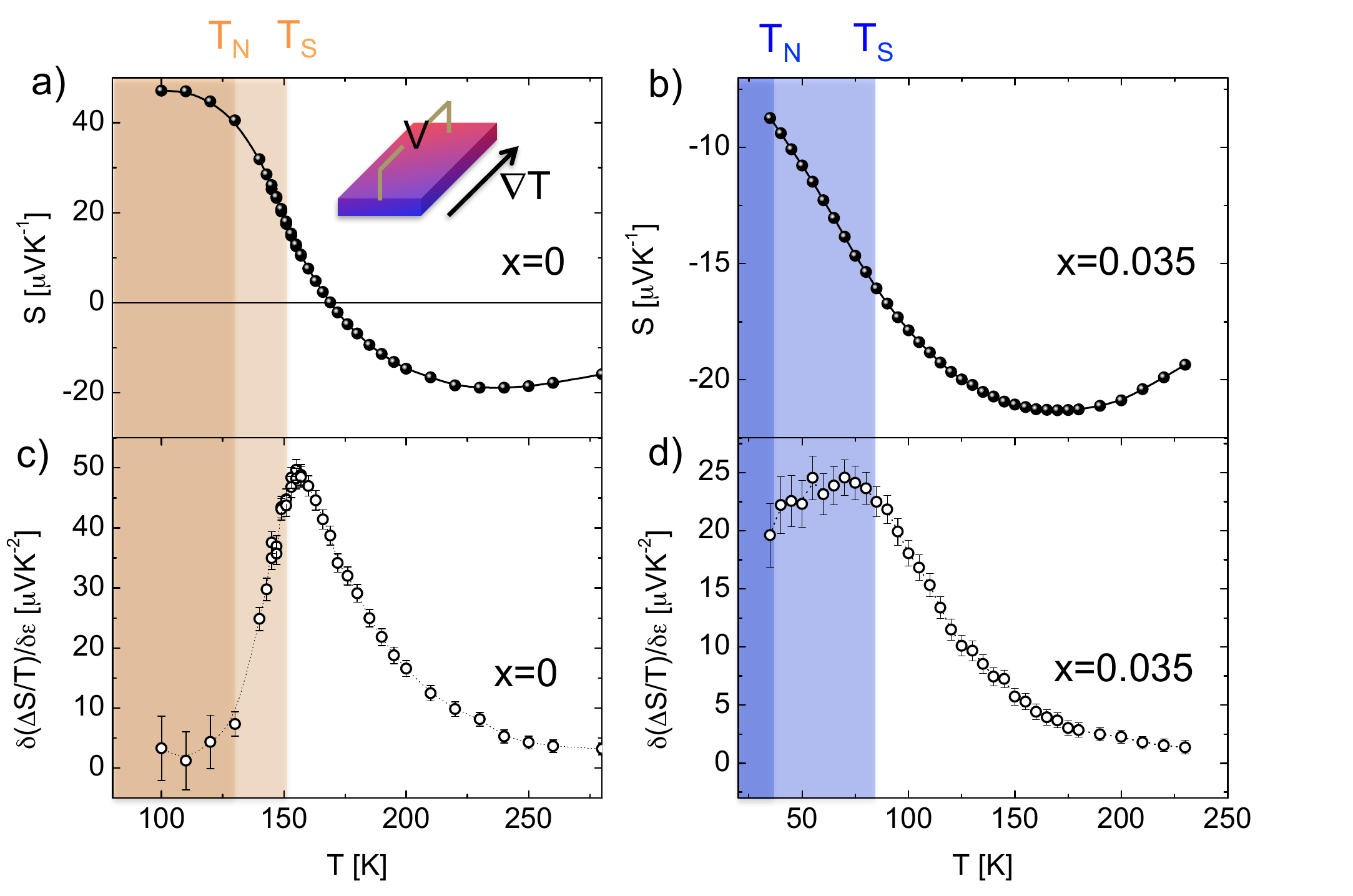}
\caption{\label{Figure} a) Temperature-dependence of the Seebeck coefficient $S$ of the LaFe$_{1-x}$Co$_x$AsO compounds with x=0. b) Temperature-dependence of the strain derivative of the $T$-normalized Seebeck coefficient  $\delta(\Delta S/T)/\delta \epsilon$ x=0 compound. c) Temperature-dependence of the Seebeck coefficient $S$ of the LaFe$_{1-x}$Co$_x$AsO compounds with x=0.035. b) Temperature-dependence of the strain derivative of the $T$-normalized Seebeck coefficient  $\delta(\Delta S/T)/\delta \epsilon$ x=0.035 compound. The dark-orange and the dark-blue areas indicate the regions where the long range structural and magnetic ordered states are both established.}
\end{figure*}

\begin{figure*}[!t]
\includegraphics[width=\textwidth]{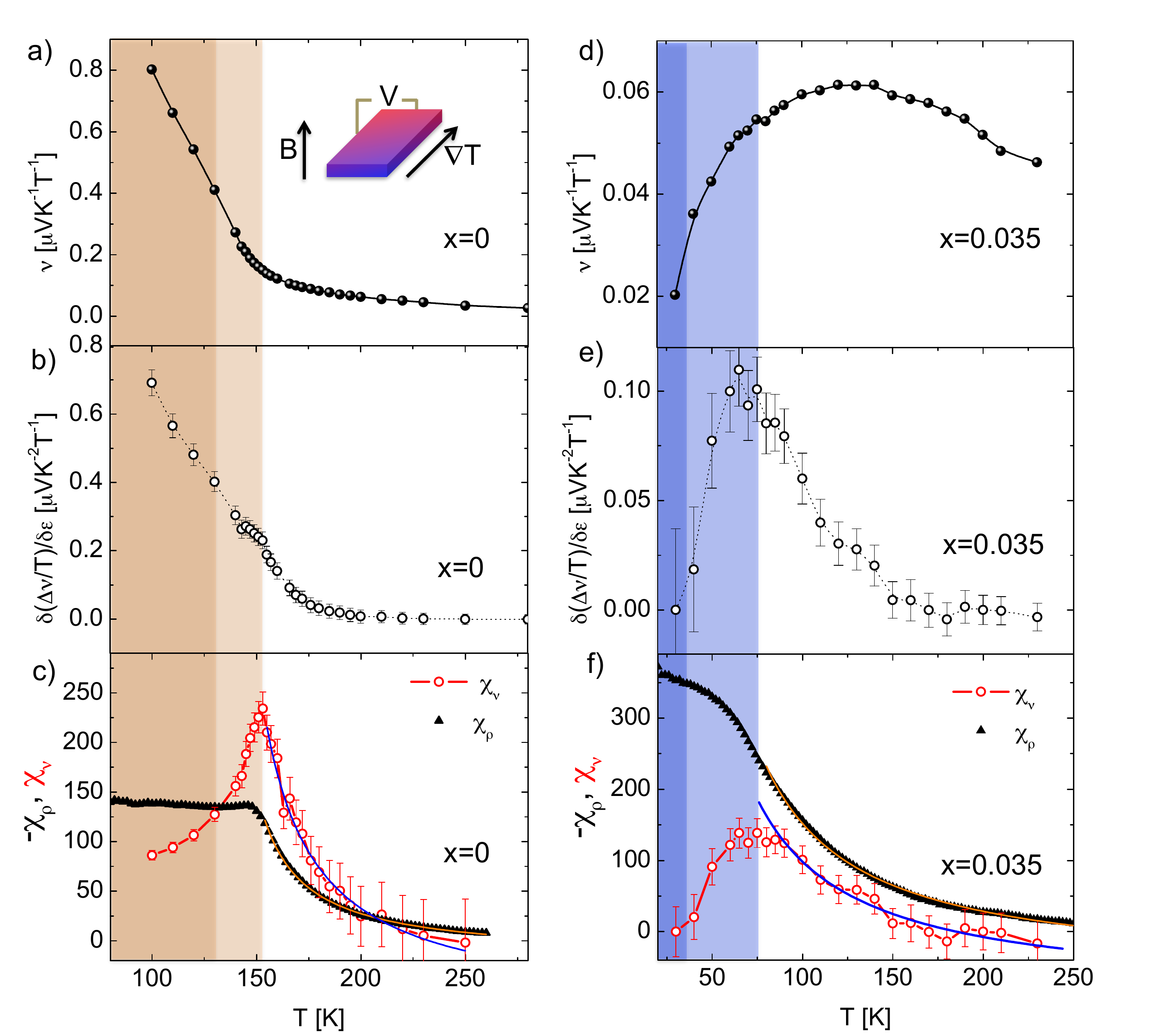}
\caption{\label{Figure} a) Temperature-dependence of the Nernst coefficient $\nu$ of the LaFe$_{1-x}$Co$_x$AsO compounds with x=0. b) Temperature-dependence of the strain derivative of the $T$-normalized Nernst coefficient  $\delta(\Delta \nu/T)/\delta \epsilon$ x=0 compound. c) Temperature-dependence of $\chi_\nu$ (red circles with red line) and -$\chi_\rho$ (black triangles) of the x=0 compound. d) Temperature-dependence of the Nernst coefficient $\nu$ of the LaFe$_{1-x}$Co$_x$AsO compounds with x=0.035. e) Temperature-dependence of the strain derivative of the $T$-normalized Nernst coefficient  $\delta(\Delta \nu/T)/\delta \epsilon$ x=0.035 compound.  f) Temperature-dependence of $\chi_\nu$ (red circles with red line) and -$\chi_\rho$ (black triangles) of the x=0.035 compound. The blue (orange) solid lines in figures c) and f) represent the Curie-Weiss fit for $T>T_S$ of $\chi_\nu$ (-$\chi_\rho$). The dark-orange and the dark-blue areas indicate the regions where the long range structural and magnetic ordered states are both established.}
\end{figure*}

\begin{figure*}[!t]
\includegraphics[width=\textwidth]{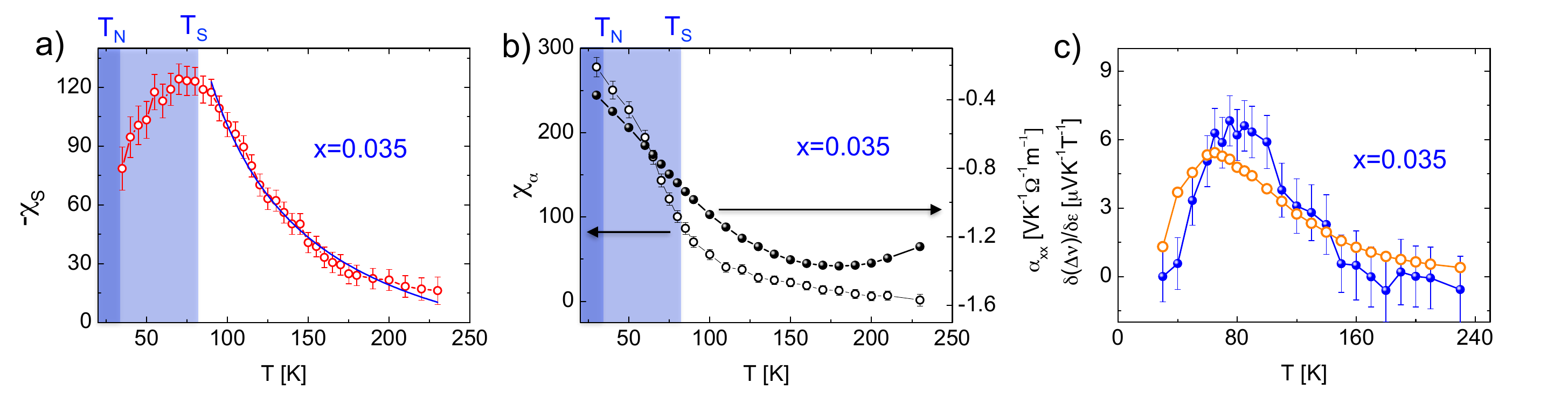}
\caption{\label{Figure} a) Temperature-dependence of -$\chi_S$ (red circles with solid line) for the x=0.035 compound. The blue solid line is the Curie-Weiss fit. b) Temperature-dependence of $\chi_{\alpha}$ (empty circles with solid line) and $\alpha_{xx}$ (black circles with solid line) for the x=0.035 compound. c) Temperature-dependence of experimental $\delta (\Delta \nu)/\delta \epsilon$ (blue circles with blue solid line) and calculated $\delta (\Delta \nu)/\delta \epsilon$ (orange empty circles with solid line) for the x=0.035 compound}
\end{figure*}

\end{document}